\documentclass[]{aastex631}
\usepackage{amsmath}
\usepackage{verbatim}
\usepackage{booktabs}
\usepackage{xcolor}

\shortauthors{Gupta et al.}

\graphicspath{{./}{figures/}}

\def\chandra {{\it Chandra}~}

\def\xmm {{\it XMM-Newton}~}

\def\suzaku {{\it Suzaku}~}

\begin{document}
\title{Thermal Structure and Chemical Enrichment of the North and South Polar Spurs: Supersolar N/O and Ne/O in the X-ray Plasma}


\correspondingauthor{Anjali Gupta}
\email{agupta1@cscc.edu}

\author{Anjali Gupta}
\affiliation{Columbus State Community College, 550 E Spring St., Columbus, OH 43215, USA} 

\author{Smita Mathur}
\affiliation{Department of Astronomy, The Ohio State University, 140 West 18th Avenue, Columbus, OH 43210, USA} 
\affiliation{Center for Cosmology and Astroparticle Physics, 191 West Woodruff Avenue, Columbus, OH 43210, USA}

\author{Joshua Kingsbury}
\affiliation{Physics Department, University of Miami, Coral Gables, FL 33124, USA } 

\author{Anthony Taylor}
\affiliation{Columbus State Community College, 550 E Spring St., Columbus, OH 43215, USA} 

\author{Sanskriti Das}
\altaffiliation{Hubble Fellow}
\affiliation{Kavli Institute for Particle Astrophysics \& Cosmology, Stanford University, 452 Lomita Mall, Stanford, CA 94305, USA}

\author{Joy Bhattacharya}
\affiliation{Department of Astronomy, The Ohio State University, 140 West 18th Avenue, Columbus, OH 43210, USA} 

\author{Manami Roy}
\affiliation{Department of Astronomy, The Ohio State University, 140 West 18th Avenue, Columbus, OH 43210, USA} 
\affiliation{Center for Cosmology and Astroparticle Physics, 191 West Woodruff Avenue, Columbus, OH 43210, USA}

\author{Yair Krongold}
\affiliation{Instituto de Astronomia, Universidad Nacional Autonoma de Mexico, 04510 Mexico City, Mexico}

\begin{abstract}
The North Polar Spur (NPS) is a prominent diffuse X-ray feature whose origin has remained uncertain for decades. Using a uniform analysis of archival \textit{Suzaku} and \textit{XMM--Newton} data with new \textit{Chandra} observations, we constrain its thermal and chemical properties. The NPS emission is fully absorbed by the neutral interstellar medium, demonstrating that the plasma lies beyond the Galactic disk and is not a local supernova remnant or nearby superbubble. The spectra require a two-temperature model with a warm--hot component ($kT \approx 0.2$ keV) and a hotter component ($kT = 0.4$--$0.7$ keV), with emission measures of $(41.8 \pm 4.9) \times 10^{-3}$ and $(12.9 \pm 2.2) \times 10^{-3} \mathrm{cm^{-6}~pc}$, respectively. A key result is the detection of super-solar abundance ratios in the warm--hot phase, with N/O $= 3.6 \pm 0.3$ and Ne/O $= 1.9 \pm 0.1$ solar. A \textit{Suzaku} observation of the outer South Polar Spur (SPS) shows similar absorption, temperatures, and enhanced abundances (N/O $= 2.9 \pm 0.4$, Ne/O $= 1.6 \pm 0.2$), though with lower emission measures. The similar super-solar abundance ratios suggest a common enrichment history. These properties are consistent with those measured along other sightlines through the X-ray--bright shells of the Galactic bubbles. Together, these results support that the NPS and SPS trace opposite limbs of the Galactic bubbles. The chemical properties suggest a strong contribution from stellar feedback in shaping the Galactic bubbles.

\end{abstract}

\section{Introduction}

The North Polar Spur (NPS) is a prominent arc-like X-ray and radio feature extending across the northern Galactic
sky. Since its discovery in the \textit{ROSAT} era, the location and origin of the NPS have remained one of the most debated topics in
Galactic astrophysics. Two main scenarios have been proposed: a nearby origin, linking the NPS to a local supernova
remnant or the Scorpius--Centaurus (Sco--Cen) OB association \citep[e.g.,][]{West2021, Das2020, Wolleben2007}, and a distant origin, associating it with a large-scale
structure near the Galactic center \citep[e.g.,][]{Sofue2019, Sofue2000}.

The local interpretation is supported by the NPS's strong X-ray and synchrotron emission \citep{Heiles2000}, its alignment with starlight polarization \citep[e.g.,][]{Panopoulou2021}, and its correlation with nearby H\,\textsc{i} filaments—features that could plausibly arise from a superbubble or supernova remnant at a distance of $\sim$100~pc \citep[e.g.,][]{Welsh2009}. However, an alternative view places the NPS much farther away, as part of the Milky Way's halo and potentially associated with the large-scale Galactic structures known as the \textit{Fermi Bubbles} \citep{Su2010}. The \textit{eROSITA} all-sky survey has revealed a pair of giant, X-ray–bright bubbles whose shells envelop the smaller, gamma-ray–emitting Fermi bubbles\footnote{Throughout this paper, we refer to the Fermi and eROSITA bubbles collectively as the ``Galactic bubbles" unless noted otherwise.} \citep{Predehl2020}.

In the southern sky, a similar arc-like structure first identified in radio maps—the South Polar Spur \citep[SPS;][]{Sofue2000}—has now been detected in bright X-ray emission with \textit{eROSITA} \citep{Predehl2020}. The NPS and SPS may be parts of the same structure: either a nearby supernova remnant or two giant, bipolar X-ray bubbles emerging from the Galactic center.

Despite decades of study, the thermal and chemical properties of the NPS remain uncertain. Prior X-ray analyses
using \textit{Suzaku} \citep{Miller2008, Kataoka2013, Kataoka2015, Tahara2015, Gu2016, Ursino2016, Yamamoto2022, Gupta2023}, \textit{XMM-Newton} \citep{Willingale2003, Gu2016}, and \textit{HaloSat} \citep{LaRocca2020} have reached widely differing conclusions, reporting plasma
temperatures ranging from $\sim$0.2 to 0.5~keV and inconsistent elemental abundance patterns. These discrepancies 
reflect differences in background modeling and data quality and have kept the physical nature of the NPS unresolved.

In this paper we settle the debate on the location and origin of the NPS. In \S~1.1 we summarize previous X-ray studies of the NPS, and in \S~1.2 we outline our motivation and approach for a uniform re-analysis of the NPS and SPS X-ray observations. To address the remaining discrepancies in earlier work, we also obtained a new \textit{Chandra} observation of the \textit{Suzaku} field of the NPS. \textit{Chandra} is uniquely capable of identifying the faintest point sources that \textit{Suzaku} cannot detect, allowing us to quantify and remove their contribution and thereby isolate the true diffuse emission from the NPS. The subsequent sections present the observations (\S~2), spectral analysis (\S~3), and results (\S~4). In \S~5, we discuss the interpretation of the NPS and SPS thermal and chemical structure and compare these properties with those measured in all-sky surveys.

\subsection{Previous NPS X-ray Studies with Conflicting Conclusions}

The NPS has been studied extensively in X-rays, yet past investigations have reached markedly different conclusions.
Temperatures spanning $\sim$0.2--0.5~keV and a wide range of elemental abundance patterns have been reported, with
different works favoring either a local interpretation or a distant, Galactic-scale origin. The NPS emission is extracted from
soft diffuse X-ray background (SDXB) observations, and discrepancies among previous studies mainly arise from
differing treatments of the SDXB and of the foreground--background separation. As a result, whether the NPS is a single-temperature or multi-temperature plasma, chemically enriched or depleted, in equilibrium or
non-equilibrium ionization---remain unanswered. Consequently, its location and origin remain unresolved.

Previous X-ray studies of the NPS can broadly be divided into two categories: pointed observations targeting the brightest segments of the spur, and wide-area surveys that probe the large-scale structure and extent of the NPS/Loop~I region across the sky. We first summarize results from the pointed observations of the bright NPS arc.

\citet{Willingale2003} analyzed \textit{XMM-Newton} observations of three NPS fields at slightly different latitudes. They modeled the NPS emission using a partially absorbed, single-temperature collisional ionisation equilibrium (CIE) plasma with variable O, Ne, Mg, and Fe abundances, finding a temperature of $kT \approx 0.26$~keV and depleted O and Ne. The authors also proposed that the NPS lies behind at least 50\% of the cold neutral gas in the absorbing disk.

\citet{Miller2008}, using \textit{Suzaku} data, adopted a similar modeling approach but additionally allowed the nitrogen abundance to vary. Their fits suggested a temperature of $kT \approx 0.3$~keV with depleted C, O, Ne, Mg, and Fe abundances of less than 0.5 solar, but a strongly enhanced nitrogen abundance, with $\mathrm{N/O} = 4.0 \pm 0.5$. The authors concluded that the N/O enhancement cannot be caused by the Sco-Cen OB association, but could result from a previous enrichment episode in the solar neighborhood.

\citet{Gu2016} reanalyzed both \textit{Suzaku} and \textit{XMM-Newton} data and introduced an ionized absorber component in addition to neutral absorption. They modeled the NPS with a single-phase CIE plasma at $kT \approx 0.25$~keV and abundances of 0.4--0.8 solar for N, O, Ne, Mg, and Fe, attributing the excess near 0.5~keV to ionized absorption rather than enhanced N/O. Further, they speculated that the NPS is a structure in the Galactic halo, due to which the X-ray emission is mostly absorbed by Galactic ISM in the line of sight.

Using \textit{Suzaku} shadow observations toward the molecular cloud MBM36, \citet{Ursino2016} separated local and distant contributions to the soft X-ray background along a line of sight near the NPS. Their analysis revealed two non-local thermal components with temperatures of $kT \approx 0.12$~keV and $kT \approx 0.29$~keV, with the hotter component consistent with the typical temperatures reported for the NPS emission. Based on geometrical and pressure arguments, the authors ruled out a nearby superbubble origin and suggested that the emission is more consistent with a large-scale Galactic structure associated with the NPS/Loop~I system.

Complementary to these pointed observations, several studies have investigated the NPS and surrounding Loop~I region over much larger angular scales. These works trace the thermal structure of the Galactic halo and its possible interaction with the Fermi bubbles. Using \textit{Suzaku} observations across multiple sightlines near the bubble edges, \citet{Kataoka2013} reported diffuse X-ray emission characterized by Galactic halo plasma at $kT \approx 0.3$~keV. The authors interpreted this component as weakly shocked gas associated with the expansion of the Fermi bubbles.

Subsequent studies extended this analysis to larger regions of the sky. \citet{Kataoka2015} analyzed additional \textit{Suzaku} and \textit{Swift} observations covering Galactic latitudes up to $|b| \approx 60^\circ$, finding that the Galactic halo plasma maintains an approximately uniform temperature of $kT \approx 0.3$~keV while the emission measure varies with position relative to the bubble edges. Similarly, \citet{Tahara2015} reported \textit{Suzaku} observations of large-scale structures near the bubble boundaries, including the ``N-cap'' and ``SE-claw'' regions, confirming the presence of diffuse $kT \approx 0.3$~keV halo plasma and identifying localized excess emission that may require an additional hotter component with $kT \approx 0.7$~keV.

\cite{LaRocca2020} used 14 HaloSat fields covering the entire NPS and fitted the spectra with two equilibrium thermal components: a warm phase ($kT \approx 0.087$~keV) and a hot phase ($kT \approx 0.28$~keV). Their derived distance of 0.4--0.8~kpc favored a non-local interpretation.

More recently, \cite{Yamamoto2022} suggested that the plasma in the NPS/Loop~I region may be in non-equilibrium ionization (NEI), finding higher temperatures ($kT = 0.4$--0.5~keV) and lower emission measures compared to CIE fits. Their NEI model suggests that the NPS/Loop~I structure has resulted from multiple explosions near the Galactic center.

In our recent analysis of roughly 250 \textit{Suzaku} observations exploring the Galactic bubbles and halo \citep{Gupta2023}, we found that the X-ray emission from the bubbles—including the NPS/Loop~I and the extended halo—is best described by a two-temperature model: a warm–hot phase near the Galaxy’s virial temperature ($kT \approx 0.2$~keV) and a hot phase at super-virial temperatures ($kT \approx 0.6$~keV). We also reported supersolar nitrogen (N/O) abundances across the sky and enhanced neon (Ne/O) abundances along some sightlines toward the Galactic bubbles. These results suggest that the NPS is part of the Galactic bubbles, as the temperatures of both components are consistent with those seen along other bubble sightlines.

These findings motivated a consistent re-analysis of the NPS with updated models and data, as presented in this work.

\subsection{Re-analysis of NPS X-ray Observations}

Previous X-ray studies of the NPS have produced inconsistent results due to differences in background modeling, spectral assumptions, and instrumental characteristics. To address these discrepancies, we perform a uniform re-analysis of all available high-quality NPS data, combining archival \textit{Suzaku} and \textit{XMM-Newton} observations with  new \textit{Chandra} observations. 

We apply consistent spectral modeling and background treatments to all datasets. In addition, we analyze an archival \textit{Suzaku} observation probing the outer region of the SPS to compare its thermal and chemical properties with those of the NPS and to test whether both features trace the same large-scale structure. The locations of the NPS and SPS fields analyzed in this work are shown in Figure~1.

Recent progress from shadow observations provides further motivation for this re-analysis. In particular, recent \textit{Suzaku} shadow observations \citep{Gupta2025} have eliminated any ambiguity surrounding the origin of the N\,\textsc{vii} emission line, ruling out any local (unabsorbed) contribution. These observations robustly demonstrate that the Milky Way's CGM contains nitrogen-rich plasma, with a supersolar average abundance ratio of $\mathrm{N/O} = 2.6 \pm 0.5$, and suggest that nitrogen-enhanced gas is widespread throughout the CGM. These new constraints strengthen the motivation for a uniform, high-quality re-analysis of the NPS and SPS, to determine whether their thermal and chemical properties are consistent with those measured along other sightlines through the Galactic bubbles.

\section{Observational Data and Reduction Methods}

\subsection{Chandra Observation of the North Polar Spur}
A precise determination of the NPS plasma properties requires isolating the truly diffuse emission, which is not possible with \textit{Suzaku} data alone due to its limited angular resolution. \textit{Chandra}'s ability to identify faint point sources is therefore critical for determining how much of the \textit{Suzaku} emission is intrinsic to the NPS. To identify and exclude point sources that could contaminate the diffuse X-ray emission, we obtained two \textit{Chandra}/ACIS-I observations of the \textit{Suzaku} NPS field (PI: A. Gupta; see Table~1), with a total exposure of $\sim$30~ks. With its superior angular resolution, \textit{Chandra} can reliably detect faint point sources that \textit{Suzaku} cannot resolve, resulting in a cleaner measurement of the diffuse emission.

The \textit{Chandra} data were reduced using CIAO (v4.15) with CALDB (v4.10.7). The event files were reprocessed with \texttt{chandra\_repro} to apply the latest calibrations and to filter out background flares. The two exposures were then merged. Merged images and exposure maps were generated in the 0.5--2.0~keV band, and point sources were detected using \texttt{wavdetect}, followed by visual verification to avoid spurious detections. A total of 25 point sources were identified. These source regions were excluded from the corresponding \texttt{Suzaku} extraction regions for the extended diffuse emission.

\subsection{Suzaku Observation of the North Polar Spur}
The NPS was observed with \textit{Suzaku} on 2005 October 3--4 (ObsID 100038010; PI: \textit{Suzaku~SWG}) with a total exposure of 46.1~ks 
and a clean exposure of 35.7~ks (Table~1).

The \textit{Suzaku} data were processed using \texttt{HEAsoft v6.29} following standard procedures. We used data from the 
back-illuminated XIS1 detector, which has the highest sensitivity below 1~keV. Using the tasks \texttt{xis3x3to5x5} 
and \texttt{ftmerge}, we combined events recorded in both the 3$\times$3 and 5$\times$5 editing modes. Time intervals 
with geomagnetic cut-off rigidity (COR) $<8$~GV were removed to minimize the particle background.

Using the \textit{Chandra} observation covering the \textit{Suzaku} field, we detected 25 point sources. The brightest source was excluded using a circular region of radius 2.5$'$, while the remaining sources were removed 
using circular masks of radius 1$'$ (Fig.~2). Although the \textit{Suzaku} XRT has a relatively broad point-spread function 
(half-power diameter $\sim$2$'$--2.5$'$), masking such large radii around every source would remove a 
substantial fraction of the usable field of view. A 1$'$ exclusion radius, commonly adopted in diffuse-emission studies, 
effectively removes the bright PSF core where most of the contaminating flux originates while preserving sufficient 
area for spectral analysis. Any residual contribution from the PSF wings is expected to be spectrally smooth and 
small compared to the statistical and background uncertainties in the diffuse NPS emission.
To evaluate the possible contribution from the \textit{Suzaku} PSF wings, we also generated a spectrum 
using the larger 2$'$ exclusion regions for all detected sources. The effect of using larger exclusion regions 
on the spectral results is examined in \S3. 

Redistribution matrix files (RMFs) and ancillary response files (ARFs) were 
generated with \texttt{xisrmfgen} and \texttt{xissimarfgen}, and the non--X-ray background (NXB) was estimated with 
\texttt{xisnxbgen}. Spectra were grouped to a minimum of 25 counts per bin to enable $\chi^2$ minimization.

\subsection{\textit{Suzaku} Observation of the South Polar Spur}

To explore the SPS, we searched the \textit{Suzaku} archive for observations in the region of the SPS with exposures longer than 20~ks. 
No empty fields (fields with no bright source in the field of view) were available probing the SPS; however, we identified one observation (ObsID~701052010) located in its outer region (see Table~1). 
Although the field contains an ultraluminous infrared galaxy (IRAS~19254$-$7245, also known as the \textit{SuperAntennae}) near the center, it provides a useful estimate of the diffuse X-ray emission toward the SPS. 

We excluded the central galaxy by removing a 2.5\arcmin-radius circular region centered on its position. To evaluate the possible contribution from scattered light of IRAS~19254$-$7245, we repeated the spectral extraction using several exclusion radii (2\arcmin, 2.5\arcmin, and 3\arcmin) around the galaxy. The effect of using different exclusion radii on the spectral results is examined in \S3. 

The SPS data were reduced following the same procedures described for the NPS using HEAsoft~v6.29 and standard \textit{Suzaku} screening criteria.

\subsection{\textit{XMM-Newton} Observations of the North Polar Spur}

The NPS was also observed with \textit{XMM-Newton} along six sightlines in 2001 (ObsIDs~0067340101--0067340601; PI: Willingale). 
The first three fields (I--III) were strongly affected by background flaring and were excluded from our analysis. 
We therefore focus on Fields~IV--VI (ObsIDs~0067340401--0067340601), which provide the best-quality data with clean exposures of 8439, 6939, and 9420~s, respectively (see Table~1). 

Data reduction was performed using the \textit{XMM-Newton} Extended Source Analysis Software (ESAS~v22.1.0) within the Science Analysis System (SAS~v20.0.0). 
We used data from the MOS1 detector for its better spectral resolution compared to the PN detector. We also processed the MOS2 data for the same fields following the same reduction procedures.
The \texttt{espfilt} task was applied to remove intervals affected by soft-proton flaring, resulting in the clean exposure times quoted above. 
Residual soft-proton contamination after filtering was modeled explicitly during spectral analysis. 

Point sources were detected and masked using the \texttt{cheese} task in ESAS, which identifies and excludes sources down to the field sensitivity limit. 
The final spectra, RMFs, ARFs, and non-X-ray background (NXB) components were generated using the \texttt{mosspectra} task. 
Spectra were grouped to a minimum of 25~counts per bin using \texttt{grppha} to allow $\chi^{2}$ minimization during fitting.

\section{Spectral Analysis}

The X-ray emission from the Galactic spurs must be isolated from the various components of the SDXB through
careful spectral modeling. This task is challenging because the SDXB includes emission from multiple physical
sources, such as Solar Wind Charge eXchance (SWCX), the Local Hot Bubble (LHB), the extragalactic Cosmic X-ray Background (CXB)—composed of both resolved
and unresolved point sources, and the Galactic halo. 

A typical SDXB spectrum is described using a three- to four-component model, commonly referred to as the
\textit{standard model} \citep[see][]{Gupta2025}. This model includes: (1) a foreground, unabsorbed component from
the LHB, represented by thermal plasma emission in CIE with a temperature of $kT = 0.1$~keV; (2) a foreground
component from SWCX, typically modeled with the \texttt{ACX2} charge-exchange code; (3) a background component
from the CXB, modeled as an absorbed power law; and (4) a Galactic halo component, described as an absorbed,
one-temperature CIE plasma representing the large-scale hot gas in the halo. Toward regions dominated by bright
extended structures such as the NPS and SPS, this last component is expected to represent emission from the
structures themselves rather than from the general halo.

Several studies have shown that the standard model may not fully reproduce observed SDXB spectra, and that
additional components are often required to obtain acceptable fits \citep[e.g.,][]{Das2019, Bluem2022, Bhattacharyya2023, Gupta2025}. In this work, we adopted
the standard model as our baseline and introduced extra thermal or line components as needed to achieve
statistically and physically acceptable fits for each field.

Spectral fitting was performed using the \texttt{XSPEC} package (version 12.10.1f). All thermal plasma components
were modeled in CIE using the \texttt{APEC} code (version 3.0.9; \citealt{Smith2001}), and solar relative abundances
were adopted from \citet{Anders1989}. Absorption by cold interstellar gas in the Galactic disk was modeled with the
\texttt{phabs} component in \texttt{XSPEC}. Unless otherwise noted, all quoted errors correspond to $1\sigma$
confidence intervals for a single parameter of interest.

\subsection{The North Polar Spur}

We began our analysis by fitting the \textit{Suzaku} and \textit{XMM-Newton}~MOS1 spectra of the NPS fields using the
standard SDXB model described above. For the foreground emission, we included two components: one
representing the LHB and another representing SWCX. The LHB emission was modeled with a single-temperature,
unabsorbed CIE plasma with solar abundances, with both its temperature and emission measure fixed to values
derived from the all-sky thermal emission maps of \citet{Liu2017}. The SWCX component was modeled using the
\texttt{ACX2} code \citep{Smith2014,Foster2020}, which incorporates velocity-dependent charge-exchange cross
sections from the Kronos database \citep{Mullen2016,Mullen2017,Cumbee2018}. A uniform solar-wind velocity of
450~km~s$^{-1}$ was adopted, and other ACX2 parameters—such as helium fraction, recombination type, and
elemental abundances—were fixed at their solar values. Tests with reasonable variations in solar-wind velocity and
ACX2 parameters confirmed that these assumptions have negligible impact on the modeled spectra at CCD
resolution.

The extragalactic CXB was represented by an absorbed power law with a photon index fixed at $\Gamma = 1.45$
\citep{Cappelluti2017}, while allowing its normalization to vary freely. For the NPS emission itself, we initially
adopted a single absorbed CIE plasma component with solar abundances (a VAPEC model). This baseline model
left significant residuals at both low ($\approx$0.5~keV) and high ($\approx$0.7-1.0~keV) energies, indicating that
additional components were required (Figure~3).

To improve the fits, we tested several physically motivated alternatives proposed in previous studies. First, we examined whether the NPS emission could be explained as a partially absorbed or unabsorbed local feature. For this test, we adopted a variant of the standard SDXB model in which the neutral hydrogen column density, $N_{\mathrm{H}}$, was allowed to vary freely (up to the full Galactic value listed in Table~1), rather than being fixed. This modification did not improve the fit, as the best-fit $N_{\mathrm{H}}$ remained pegged at the upper limit. Because we observe absorption from within the Galactic disk, a best-fit $N_{\mathrm{H}}$ equal to the full Galactic column indicates that the emitting plasma lies behind the bulk of the absorbing disk gas.
This rules out a very nearby ($\lesssim$100 pc) origin and is consistent with a distant, kiloparsec-scale Galactic structure such as the Galactic bubbles. However, confirmation of its precise distance would ultimately require dedicated shadow observations.

Assuming that the NPS emission is absorbed and distant, we followed the two-temperature approach
developed in our previous \textit{XMM-Newton} \citep{Das2019, Bhattacharyya2023} and \textit{Suzaku} \citep{Gupta2021,Gupta2023,Gupta2025} studies. Adding a second absorbed CIE component
significantly improved the fits, reducing $\chi^{2}$ per degree of freedom across all fields. We also allowed in the warm-hot phase the
nitrogen (N) and neon (Ne) abundances to vary, as N\,VII and Ne\,IX exhibit strong emission lines near 0.5~keV and
0.9~keV, respectively. These lines are most prominent in plasma with temperatures around 0.2~keV, where both
ions reach their peak emissivity. A narrow Gaussian component near 0.43~keV, corresponding to N\,VI emission, was included in the 
\textit{Suzaku} model to account for possible low-temperature line emission suggested by previous analyses \citep{Gupta2025}. This N\,VI feature is not constrained in the \textit{XMM-Newton} spectra due to limited sensitivity at that energy, as the 0.43~keV line lies in the regime where MOS suffers from strong low-energy 
background and degraded calibration.

All four NPS fields required both a warm-hot component ($kT \sim 0.2$~keV) and a hotter component ($kT \sim
0.4$--0.7~keV), with super-solar nitrogen abundances (N/O~$>$~1). Three fields also showed evidence for enhanced
neon (Ne/O~$>$~1). While the temperatures of the two components were consistent across the fields, the emission
measures varied among them. The best-fit model components for the NPS fields are listed in Table~2, and the best-fit
foreground components along with the CXB parameters are listed in Table~3. Figures~4 and~5 show the best-fit
spectral models for the \textit{Suzaku} Field~1 and \textit{XMM-Newton} Field~2, respectively.

To test the possible effect of the \textit{Suzaku} PSF wings from masked point sources, we repeated the spectral analysis using larger 2$'$ exclusion regions for all detected sources. The resulting temperatures, emission measures, and abundance ratios remain consistent within the statistical uncertainties, although the signal-to-noise ratio becomes slightly lower due to the reduced extraction area. This confirms that residual contamination from the \textit{Suzaku} PSF wings does not significantly affect the derived properties of the diffuse NPS emission.

We also analyzed the \textit{XMM-Newton} MOS2 spectra for the same fields. The best-fit temperatures, emission measures, and abundance ratios are consistent with those derived from the MOS1 spectra within the statistical uncertainties.

\subsection{The South Polar Spur}
We began by fitting the \textit{Suzaku} spectrum of the field probing the SPS using the standard SDXB model.  
As in the NPS case, this single-temperature model left noticeable residuals at both low and high energies.  
Adding a second thermal component and allowing the nitrogen (N) and/or neon (Ne) abundances to vary significantly improved the fit.  
Figure~6 shows the best-fit model for SPS Field~1, and Table~2 and Table~3 lists the corresponding best-fit parameters.

We also tested partially absorbed and other alternative models discussed above, but the absorbed two-temperature model with super-solar N and/or Ne abundances provided the best fit (Table~4).

We also tested a model that included the narrow Gaussian component near 0.43~keV used in the NPS analysis to represent possible N\,VI emission. In the SPS spectrum this component is not statistically required and does not improve the fit. Therefore, we adopt the simpler model without the additional Gaussian component for the SPS field. The SPS is much fainter than the NPS, resulting in a spectrum with significantly lower S/N. This is why it was not possible to detect the N\,VI emission line above the rest of the model of the SPS. This is consistent with the results of \citet{Gupta2023} in which the line was detected only in high S/N spectra of bright and/or deep fields.

The SPS temperatures derived here ($kT \approx 0.22$~keV for the 
warm--hot component and $kT \approx 0.50$~keV for the hotter plasma) are 
consistent with those measured toward the NPS within uncertainties, 
suggesting that both structures trace similar thermal phases of the 
large-scale Galactic halo environment.

To assess the possible contribution of scattered light from IRAS~19254$-$7245, we repeated the spectral analysis using several exclusion radii (2$'$, 2.5$'$, and 3$'$) around the galaxy. The resulting temperatures, emission measures, and abundance ratios remain consistent within the statistical uncertainties. This indicates that scattered light from the galaxy does not significantly affect the derived properties of the diffuse SPS emission.


\subsection{Testing Previously Proposed Spectral Models}

As summarized in \S1.1, several spectral models have been applied to the 
same \textit{Suzaku} and \textit{XMM-Newton} NPS datasets, including 
single-temperature CIE models with variable abundances 
\citep{Willingale2003, Miller2008}, ionized-absorption models 
\citep{Gu2016}, and non-equilibrium ionization models \citep{Yamamoto2022}. To evaluate these models consistently, we tested them using the same foreground SDXB modeling.

Allowing O, Ne, Mg, and Fe abundances to vary, following \citet{Willingale2003}, 
improves the fit relative to the basic standard model but still leaves 
significant residuals near 0.5~keV and 0.8--1.0~keV. Including nitrogen as a 
free parameter, as in \citet{Miller2008}, provides further improvement but 
again fails to account for the high-energy residuals. The ionized-absorption 
model of \citet{Gu2016} reduces the 0.5~keV excess but does not match the 
overall spectral shape as well as our baseline model, even with SWCX already 
included. NEI models also do not provide statistically acceptable fits.

In all cases, the absorbed two-temperature model with variable N/O and Ne/O 
ratios yields the best overall fit (Table~4), consistent with the presence of 
both a warm--hot and a hotter plasma component.

\section{Results}
The X-ray emission spectra of both the NPS and SPS are well described by an absorbed two-temperature model, with both regions exhibiting evidence of super-solar nitrogen (N/O) and neon (Ne/O) abundances.  
The NPS exhibits an average N/O ratio of $\mathrm{3.6 \pm 0.3}$ and an average Ne/O ratio of $\mathrm{1.9 \pm 0.1}$, indicating enhanced abundances in most fields.  
For the SPS, we find an N/O ratio of $\mathrm{2.9 \pm 0.4}$ and a Ne/O ratio of $\mathrm{1.6 \pm 0.2}$.  
The \textit{Suzaku} NPS field also requires a Gaussian emission line corresponding to N~VI, suggesting the presence of a low-temperature plasma component.

The temperature of the warm-hot phase in both regions is consistent with the Galaxy’s virial temperature, around $kT \approx 0.2$~keV. For the NPS, the average emission measures (EMs, $EM = \int n_e n_H dl$) are $\mathrm{EM_{warm-hot} = 41.8 \pm 4.9\times 10^{-3}~\mathrm{cm^{-6}~pc}}$ and $\mathrm{EM_{hot} = 12.9 \pm 2.2 \times 10^{-3}~\mathrm{cm^{-6}~pc}}$.  For the SPS, the corresponding values are $\mathrm{EM_{warm-hot} = 18.4 \pm 1.0\times 10^{-3}~\mathrm{cm^{-6}~pc}}$ and $\mathrm{EM_{hot} = 4.1 \pm 0.5\times 10^{-3}~\mathrm{cm^{-6}~pc}}$.  In both components, the NPS shows systematically higher EMs than the SPS, indicating a denser or more extended emitting plasma in the observed fields.

\section{Discussion \& Conclusion}

\subsection{Comparison with Previous Measurements}

To provide a quantitative comparison with previous works, Figure~7 
summarizes the plasma temperatures reported in major X-ray studies of the 
NPS/Loop~I region. The comparison includes both single-temperature and 
multi-temperature spectral analyses based on \textit{Suzaku}, 
\textit{XMM-Newton}, and HaloSat observations.

Warm--hot components reported in previous work typically fall in the range 
$kT \approx 0.25$--$0.30$~keV when modeled with a single thermal plasma 
(e.g., \citealt{Willingale2003}; \citealt{Miller2008}; 
\citealt{Kataoka2013, Kataoka2015}; \citealt{Tahara2015}). 
Our measurements for the NPS and SPS fields instead are near 
$kT \approx 0.20$--$0.22$~keV, closer to the characteristic temperature of the 
extended Galactic halo.

The hotter thermal component varies more between sightlines. In our analysis 
the hot plasma temperature is $kT = 0.491 \pm 0.050$~keV for the NPS field and 
$kT = 0.495 \pm 0.041$~keV for the SPS field. For comparison, 
\citet{Gupta2023} report an average hot-component temperature of 
$kT = 0.741 \pm 0.018$~keV toward the Galactic bubbles region. 
Temperatures measured along their bubble sightlines span a broader range of 
roughly $kT \approx 0.4$--$1.1$~keV.

For reference, the hatched horizontal bands in Figure~7 show the 
characteristic temperatures of the extended Galactic halo measured by 
\citet{Gupta2023}. The lower band corresponds to the warm--hot halo 
component ($kT \approx 0.20$~keV), while the upper band marks the hotter 
halo phase ($kT \approx 0.84$~keV). The comparison shows that the warm--hot 
temperatures derived for the NPS and SPS are consistent with the extended 
halo value, while the hotter component lies below the typical hot halo 
temperature but remains within the broader range reported toward the 
Galactic bubbles in previous studies.

Most previous X-ray studies of the NPS assumed solar relative 
abundances or fixed abundance ratios, making direct comparisons of elemental 
ratios difficult. For example, \citet{Kataoka2013,Kataoka2015,Tahara2015} 
adopted a metallicity of $\sim0.2$ solar while keeping the relative 
abundances fixed to solar values.

Using \textit{Suzaku} observations, \citet{Miller2008} measured a 
strongly enhanced nitrogen abundance with $N/O \approx 4.0$ times the solar 
value. Our measured $N/O$ ratios are consistent with this earlier result.

Several studies also reported enhanced neon relative to oxygen. 
\citet{Willingale2003} allowed several metal abundances to vary and reported 
values of $O = 0.32\pm0.02$ and $Ne = 0.44\pm0.04$ (in solar units), 
corresponding to $Ne/O \approx 1.38$.

A later reanalysis by \citet{Gu2016} obtained oxygen abundances of 
$\sim0.2$--$0.3$ solar and neon abundances of $\sim0.3$--$0.6$ solar, 
implying $Ne/O$ ratios of order $\sim1$--$~2$ depending on the spectral 
model.

Finally, \citet{Ursino2016} reported $Ne/O = 1.7\pm0.3$ from shadowing 
observations probing a region near the NPS. Overall, these measurements 
indicate super-solar $Ne/O$ ratios, broadly consistent with the values 
derived in this work. This comparison demonstrates that both $N/O$ and 
$Ne/O$ enhancements inferred in this work are consistent with previous 
measurements of the NPS X-ray plasma and that similar abundance patterns 
are also observed toward the SPS.

\subsection{Interpretation of the  Thermal and Chemical Properties of NPS and SPS}
Previous X-ray studies have reported a wide range of temperatures, abundances, and interpretations for the NPS,
motivating our uniform re-analysis using consistent models.
We tested a wide range of spectral models to fit the \textit{Suzaku} and \textit{XMM-Newton} observations of both the NPS and SPS to assess the possible location of the emitting
plasma. The unabsorbed and partially absorbed thermal models models failed to reproduce the observed spectra. In contrast, our fits show that the NPS emission is
fully absorbed by the neutral interstellar medium (\S3.1), requiring the plasma to lie beyond the absorbing
disk. This rules out a local supernova remnant or nearby superbubble origin and instead indicates that the NPS is a
distant, Galactic-scale feature consistent with the morphology of the large X-ray bubbles.

We next explored several physically motivated spectral models for the NPS and SPS emission, including:
(i) single-temperature models with variable abundances;
(ii) non-equilibrium ionization models;
(iii) single-temperature model with ionized absorption; and
(iv) two-temperature models with variable N/O and Ne/O using neutral absorption.
Among these, the two-temperature models provide the best fits. 


We therefore conclude that the X-ray emission from both the NPS and SPS is best described by a two-temperature
model with enhanced nitrogen and neon abundances, and that both structures trace chemically enriched plasma
located beyond the Galactic disk.

\subsection{Comparison with All-Sky Surveys}

Our results are broadly consistent with recent studies of X-ray emission from the Galactic halo and the Galactic bubbles, which also favor a two-temperature model with variable abundances.  
\citet{Gupta2023} analyzed \textit{Suzaku} observations of sightlines probing the Galactic bubbles (Galactic longitudes $300^{\circ} < l < 60^{\circ}$) and the surrounding extended halo ($60^{\circ} < l < 300^{\circ}$).  We found that the X-ray emission from the shells of the Galactic bubbles is best described by a two-temperature thermal model, with one component near the Galaxy’s virial temperature and another at super-virial temperatures.  The temperatures of both components were found to be similar inside and outside the shells, though the EMs were significantly higher within the shells.  
Additionally, toward the ten (out of 150) Galactic bubble sightlines, the best-fit models required super-solar abundance ratios of $(\mathrm{Ne/O}) = 2.1 \pm 0.2$ solar. 
{\bf The models also required an overabundance of nitrogen, with an average $(\mathrm{N/O}) = 4.2 \pm 0.2$ solar in the warm–hot phase, both within and around the bubbles.}
Thus, the thermal parameters of the NPS and the SPS closely match those observed in other bubble sightlines.  Therefore, they are likely parts of the Galactic bubbles.  



\section{acknowledgments}
We gratefully acknowledge support from NASA ADAP grants \textbf{80NSSC24K0626} and \textbf{80NSSC22K0480}, awarded to AG. 
AG also acknowledges support from the National Aeronautics and Space Administration through Chandra Award Numbers 
\textbf{GO3-24126X} and \textbf{GO4-25092X}, issued by the Chandra X-ray Center, which is operated by the Smithsonian 
Astrophysical Observatory for NASA under contract NAS8-03060. 
SM acknowledges support from Chandra Award Number \textbf{AR0-23014X} and from NASA ADAP grant \textbf{80NSSC22K1121}. 
SD acknowledges support from the NASA Hubble Fellowship and the KIPAC Fellowship of the Kavli Institute for Particle 
Astrophysics and Cosmology, Stanford University. 
YK acknowledges support from grant \textbf{PAPIIT-UNAM IN102023}.

\subsection{Data Availability}
This work makes use of data obtained from the \textit{Chandra X-ray Observatory}. The datasets used in this paper are available in the Chandra Data Collection \dataset[DOI: 10.25574/cdc.558]{https://doi.org/10.25574/cdc.558}.

\newpage
\begin{figure}
\includegraphics[scale=0.5]{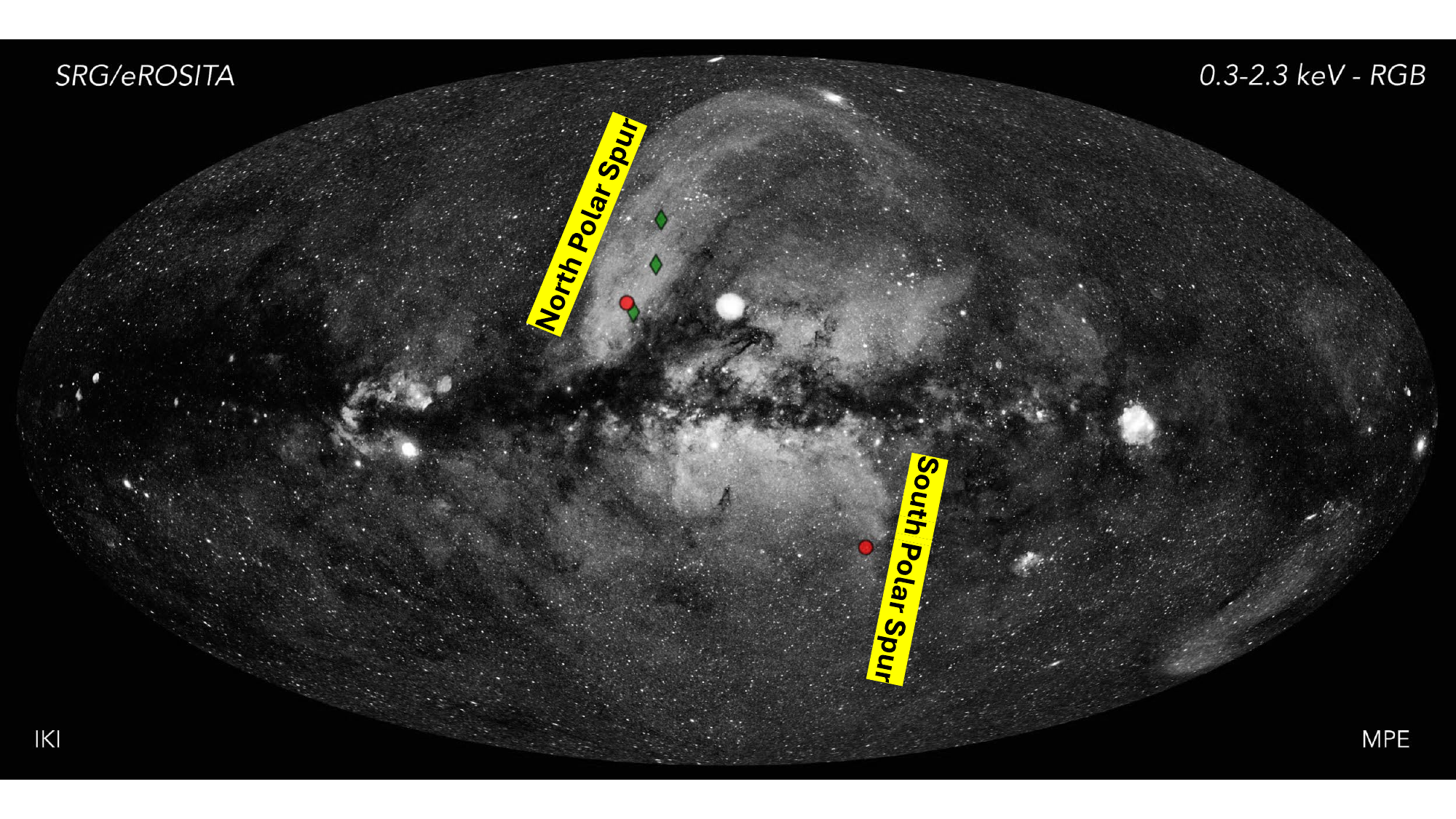}
\caption{All-sky X-ray map from \textit{eROSITA} (0.3--2.3 keV) showing large-scale Galactic emission. The NPS and SPS are labeled in yellow. The red circle marks the fields analyzed in this work: for the NPS, it denotes co-spatial \textit{Suzaku} and \textit{Chandra} observations, and for the SPS it indicates the \textit{Suzaku}-only field. Green diamonds mark the \textit{XMM-Newton} NPS fields studied here. The background image is from the SRG/\textit{eROSITA} all-sky survey (Credit: Jeremy Sanders, Hermann Brunner and the eSASS team (MPE); Eugene Churazov and Marat Gilfanov, on behalf of IKI). }
\end{figure}

\begin{figure}
\includegraphics[scale=0.65]{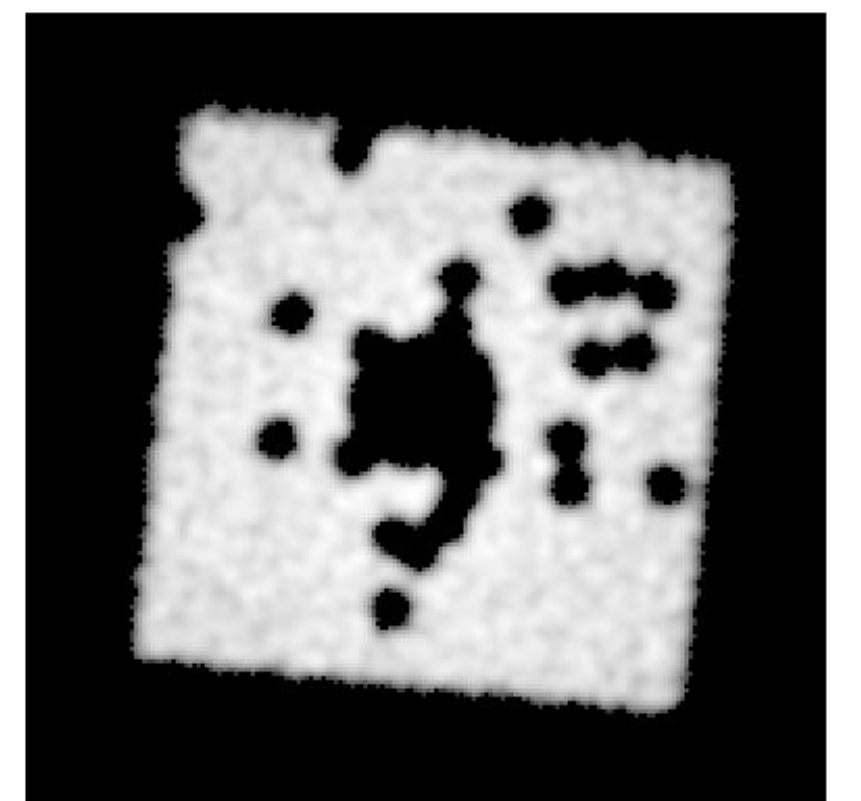}
\vspace*{-0.3 cm}
\caption{\textit{Suzaku}/XIS1 image of the NPS field after removal of point sources identified using \textit{Chandra} observations. The masked regions ($2.5^\prime$ radius for the brightest source and $1^\prime$ for the remaining sources) are excluded to isolate the diffuse emission used for spectral analysis.}
\end{figure}

\begin{figure}
\includegraphics[scale=0.8]{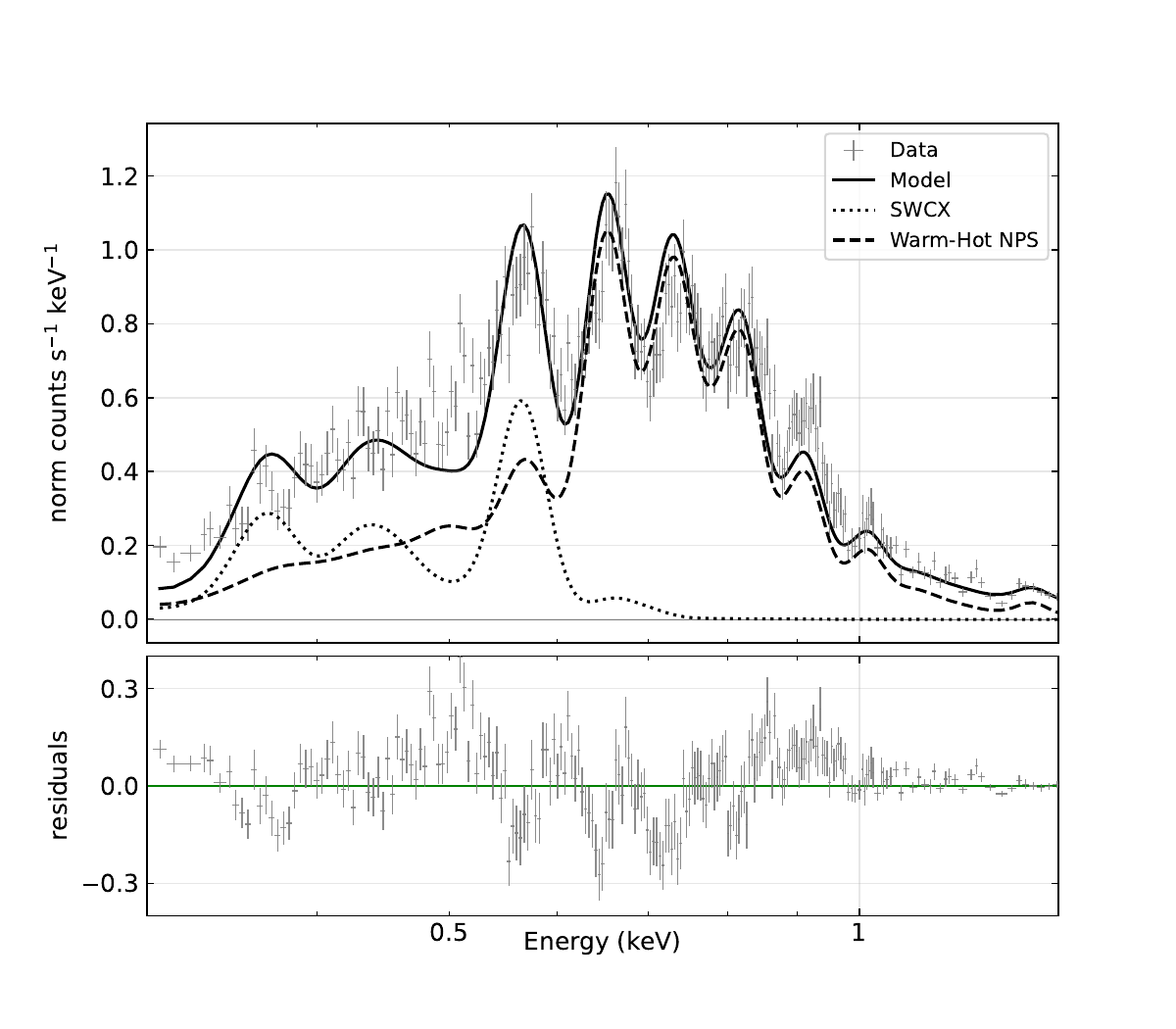}
\vspace*{-0.3 cm}
\caption{\textit{Suzaku}/XIS1 spectrum of the NPS\_field~1 fitted with the SDXB standard model. The residuals clearly show excess emission at both lower and higher energies. For clarity, the LHB and CXB components are not displayed.}
\end{figure}

\clearpage

\begin{figure}
\includegraphics[scale=0.8]{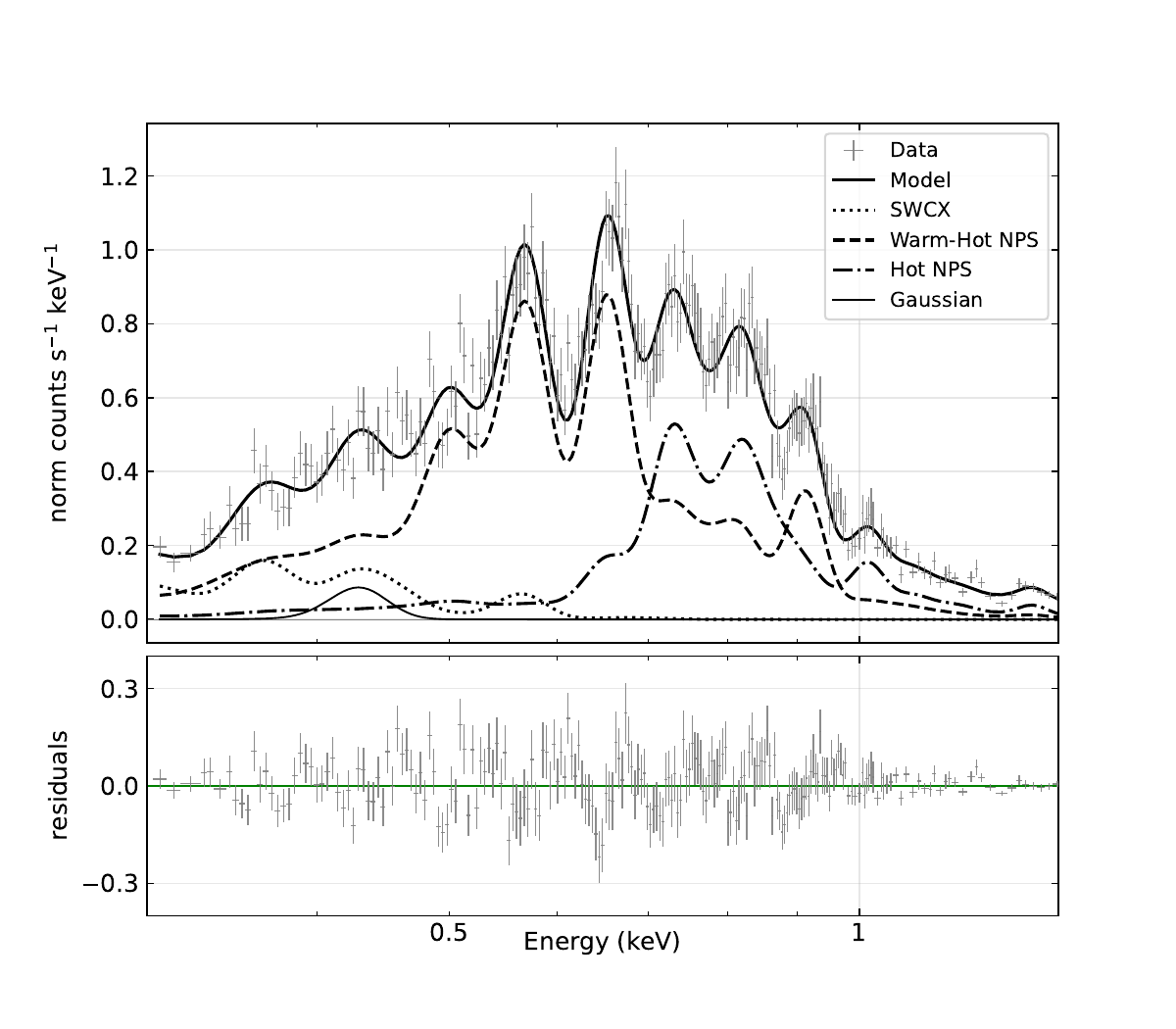}
\caption{\textit{Suzaku}/XIS1 spectrum of the NPS\_field~1 fitted with the absorbed two-temperature model allowing variable N/O and Ne/O. The dashed and dash-dotted curves show the warm–hot enriched and hot NPS components, respectively. The thin curve marks the narrow Gaussian line representing \ion{N}{6} at 0.43 keV. The dotted curve indicates the foreground SWCX emission. For clarity, the LHB, and the CXB, components are not shown. }
\end{figure}

\clearpage

\clearpage

\begin{figure}
\includegraphics[scale=0.8]{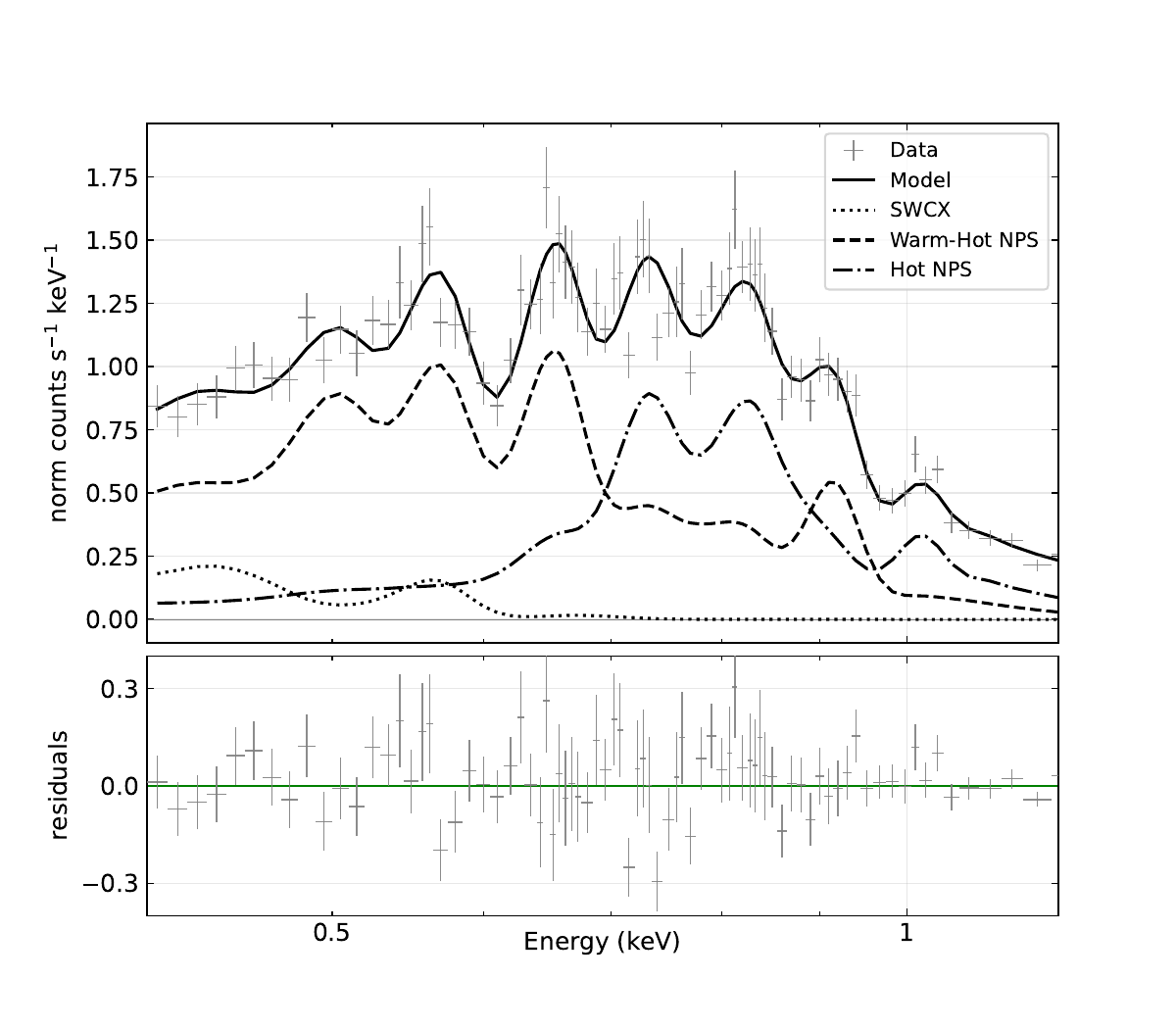}
\caption{ \textit{XMM–Newton}/MOS1 spectrum of the NPS\_field~2, fitted with an absorbed two-temperature model with variable N/O and Ne/O abundances. The warm–hot enriched and hotter NPS components are indicated by the dashed and dash-dotted curves, respectively, and the dotted curve represents the foreground SWCX contribution. The LHB, CXB, and proton-induced background components are omitted for visual clarity. }
\end{figure}

\clearpage

\begin{figure}
\includegraphics[scale=0.8]{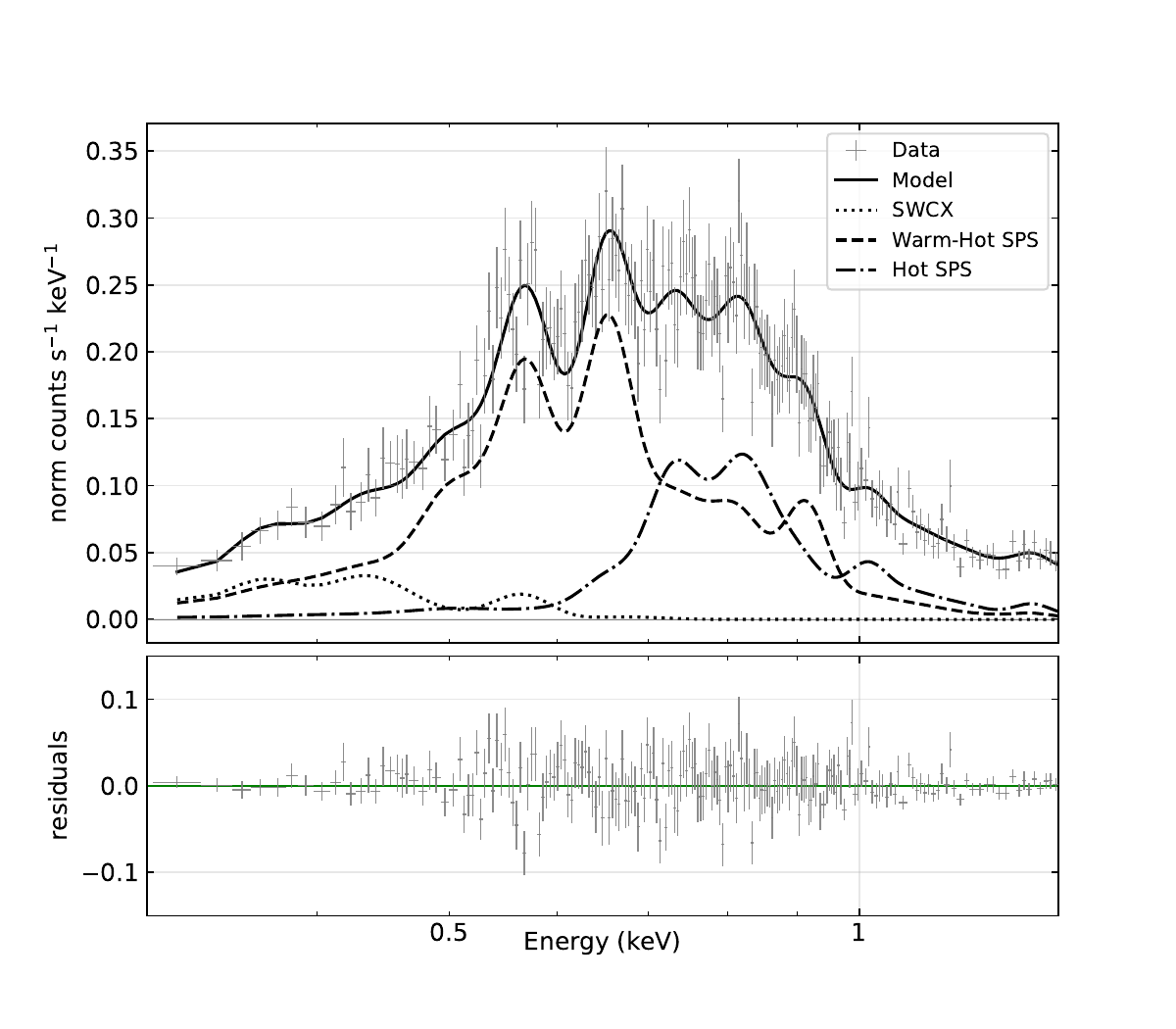}
\caption{\textit{Suzaku}/XIS1 spectrum of the SPS fitted with an absorbed two-temperature model in which the N/O and Ne/O abundances are allowed to vary. The dashed and dash-dotted curves represent the warm-hot and hot SPS components, respectively, while the dotted curve shows the contribution from foreground SWCX. For clarity, the LHB and CXB components are not displayed. }
\end{figure}

\begin{figure}
\centering
\includegraphics[width=0.95\linewidth]{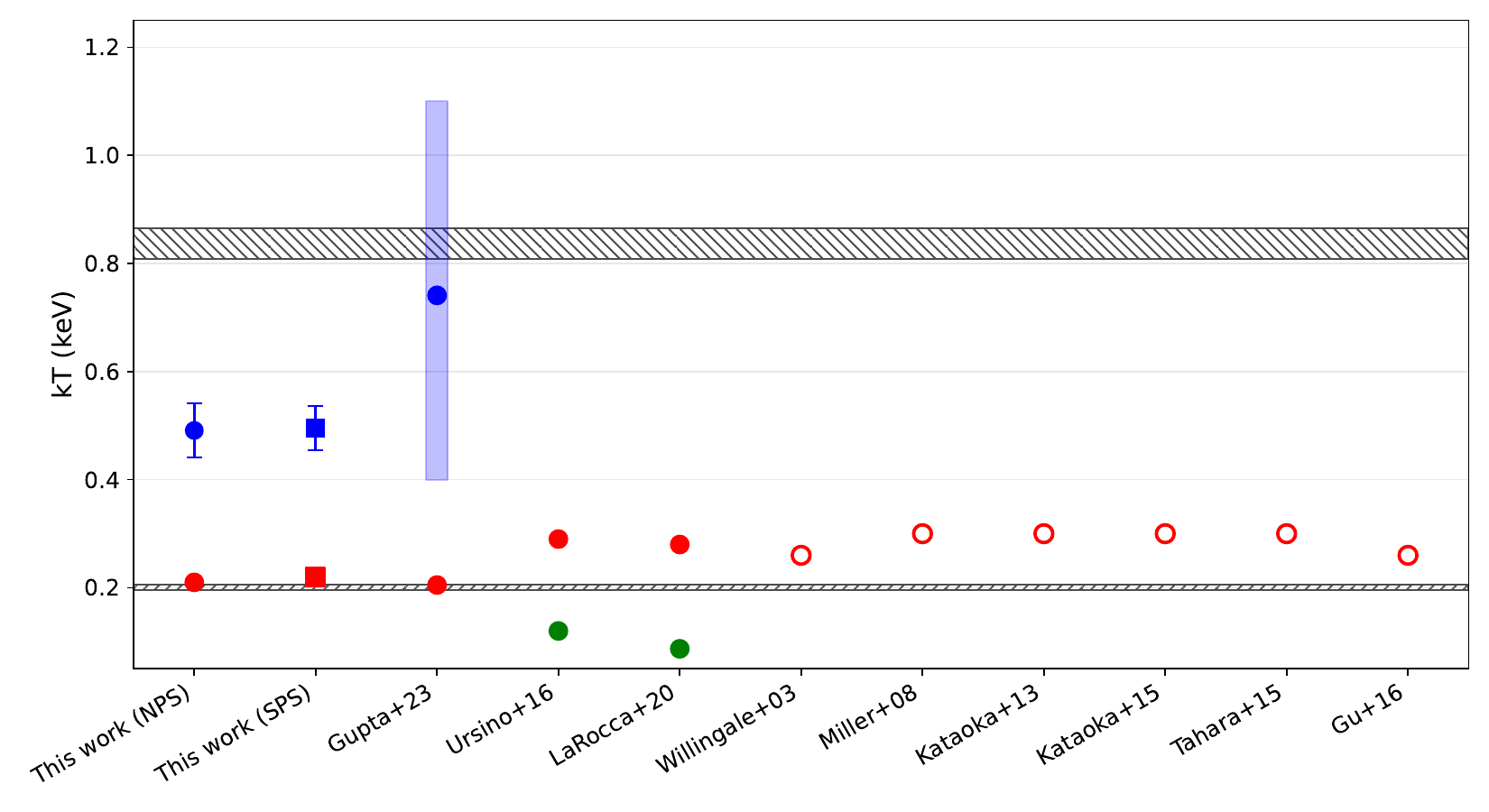}
\caption{Comparison of plasma temperatures reported in this work and 
previous X-ray studies of the NPS/Loop~I and SPS regions. 
Red symbols denote the warm--hot plasma component, blue symbols the hotter 
thermal component, and green symbols cooler halo components reported in some studies. Circles indicate measurements toward the NPS, while squares correspond to the SPS field analyzed in this work. 
The vertical blue band indicates the range of hot-component temperatures 
($kT \approx 0.4$--$1.1$~keV) measured along Galactic-bubbles sightlines in 
\citet{Gupta2023}. Hatched horizontal bands mark the temperatures of the 
warm--hot and hot phases of the extended Galactic halo.}
\end{figure}

\newpage

\begin{deluxetable*}{lccccccc}
\tabletypesize{\scriptsize}
\tablewidth{0pt} 
\tablenum{1}
\tablecaption{Observations Log \label{tab:deluxesplit}}
\tablehead{
\colhead{Target} & \colhead{Observatory} &\colhead{ObsID}  & \colhead{Start Date} & \colhead{\it l}         & \colhead{\it b}         &  \colhead{Exposure} & \colhead{$\rm N_{H}$}\\
\colhead{}    & \colhead{}          & \colhead{}    &   \colhead{}     & \colhead{($ ^{\circ}$)} & \colhead{($ ^{\circ}$)} &  \colhead{(ks)}     & \colhead{$\rm 10^{20}~cm^{-2}$} }
\startdata 
  &    &     \multicolumn{3}{c}{North Polar Spur}   &   \\
\hline
{NPS\_field~1}    & \suzaku &$100038010$ & $\rm Oct 3, 2005 $ &  $26.83$  &  $21.95$  &  $35.7$  & $5.5$  \\
{NPS\_field~1}    & \chandra &$28129$ & $\rm May 28, 2025 $ &  $26.84$  &  $21.96$  &  $14.9$  & $5.5$  \\
{NPS\_field~1}    & \chandra &$30944$ & $\rm May 29, 2025 $ &  $26.84$  &  $21.96$  &  $12.9$  & $5.5$  \\
{NPS\_field~2} & \xmm    &$67340401$  & $\rm Feb 28, 2001$ &  $25.0$  &  $20.0$  &  $8.4$  & $7.8$  \\
{NPS\_field~3}  & \xmm    &$67340501$  & $\rm Feb 28, 2001$ &  $20.0$  &  $30.0$  &  $6.9$  & $5.0$  \\
{NPS\_field~4} & \xmm    &$67340601$  & $\rm Feb 28, 2001$ &  $20.0$  &  $40.0$  &  $9.4$  & $3.5$  \\
\hline
  &    &     \multicolumn{3}{c}{South Polar Spur}   &   \\
\hline
{SPS\_Field1} & \suzaku & $701052010$ & $\rm Nov 2, 2006 $ & $322.50$  &  $-28.76$  &  $88.61$  &   $6.0$  \\
\enddata
\end{deluxetable*}

\begin{deluxetable*}{lccccccccccc}
\tabletypesize{\scriptsize}
\tablewidth{0pt} 
\tablenum{2}
\tablecaption{Model Parameters for the Galactic (NPS and SPS) Emission}
\tablehead{
\colhead{Target}  & \multicolumn{4}{c}{\bf Warm-Hot Component} &   &  \multicolumn{2}{c}{\bf Hot Component}        &     &  \colhead{$\chi^{2}/d.o.f.$}\\
\cline{2-5}
\cline{7-8}
\colhead{Model}  & \colhead{kT$_{1}^{a}$} & \colhead{N}   &\colhead{Ne} & \colhead{EM$_{1}^{b}$} &  &\colhead{kT$_{2}^{a}$} &  \colhead{EM$_{2}^{b}$} & & \colhead{}
} 
\rotate
\startdata 
{\bf NPS\_Field~1$^{*}$}       & $0.210\pm0.003$ &$3.7\pm0.2$ &$2.1\pm0.2$  & $46.0\pm1.3$ & & $0.507\pm0.025$& $13.7\pm1.0$   && $407.70/336$\\
{\bf NPS\_Field~2}     & $0.205\pm0.029$ &$3.8\pm1.0$ &$2.1\pm0.4$  & $39.6\pm10.0$ & & $0.480\pm0.039$& $13.9\pm3.5$   && $416.16/396$\\
{\bf NPS\_Field~3}       & $0.204\pm0.028$ &$3.3\pm1.1$ &$1.8\pm0.3$  & $33.9\pm8.4$ & & $0.465\pm0.046$& $8.6\pm2.7$   && $352.66/366$\\
{\bf NPS\_Field~4}      & $0.220\pm0.063$ &$3.9\pm1.1$ &$1.3\pm0.3$  & $23.4\pm3.2$ & & $0.490\pm0.060$& $6.1\pm2.6$   && $472.77/406$\\
{\bf SPS\_Field~1}       & $0.206\pm0.005$ &$2.9\pm0.4$ &$1.6\pm0.2$  & $18.4\pm1.0$ & & $0.495\pm0.041$& $4.1\pm0.5$   && $461.14/501$\\
\enddata
\tablecomments{\\
$^{a}$ Temperature is in units of keV.\\
$^{b}$ EM is in units of $10^{-3}~ cm^{-6}~ pc$\\
$^{*}$ The \suzaku NPS\_Field~1 also required a Gaussian at $E = 0.43~\mathrm{keV}$ (frozen), with line intensities of $4.2 \pm 1.5$ L.U. ($\mathrm{photons~s^{-1}~cm^{-2}~sr^{-1}}$).
}
\end{deluxetable*}

\begin{deluxetable*}{lcccccc}
\tabletypesize{\scriptsize}
\tablewidth{0pt} 
\tablenum{3}
\tablecaption{Model Parameters for the Foreground (LHB+SWCX) and the CXB components.}
\tablehead{
\colhead{Target} & \colhead{kT$_{LHB}^{a}$} & \colhead{EM$_{LHB}^{a}$} & \colhead{kT$_{SWCX}^{b}$} & \colhead{N$_{SWCX}^{b}$} & & \colhead{$\rm Norm_{PL}^{c}$}
}
\startdata 
{\bf NPS\_Field1} & $0.09$  &  $1.5$ & $0.075\pm0.004$ & $10.6\pm3.4$ & & $13.0\pm0.5$\\
{\bf NPS\_Field2} & $0.09$  &  $1.0$ & $0.096\pm0.063$ & $1.8\pm5.4$ & & $14.0\pm0.9$\\
{\bf NPS\_Field3} & $0.09$  & $1.2$ & $0.101\pm0.033$ & $2.2\pm2.4$ & & $6.8\pm0.8$\\
{\bf NPS\_Field4} & $0.09$  & $1.3$ & $0.097\pm0.039$ & $2.0\pm3.5$  & & $8.8\pm0.8$\\
{\bf SPS\_Field1} & $0.10$  & $3.0$  & $0.074\pm0.007$ & $6.7\pm5.4$ & & $10.3\pm0.3$\\
\enddata
\tablecomments{\\
$^{a}$ The LHB APEC model temperature (in units of keV) and EM (in units of $\rm 10^{-3}~cm^{-6}~pc$) fixed based on values derived from the thermal emission maps of Liu et al. (2017)\\
$^{b}$ The SWCX ACX2 model temperature (in units of keV) and Normalization (in units of $\rm 10^{-3}~cm^{-5}~sr^{-1}$).\\
$^{c}$ Normalization of the power-law model in the units of $\rm photons~keV^{-1}~s^{-1}~sr^{-1}~cm^{-2}$.\\
}
\end{deluxetable*}

\newpage
\begin{deluxetable*}{lccccc}
\tabletypesize{\scriptsize}
\tablewidth{0pt} 
\tablenum{4}
\tablecaption{Fit Statistics ($\chi^{2}/d.o.f$) for Different Models for the NPS and SPS}
\tablehead{
\colhead{Model/Target} & \colhead{NPS\_Field1} & \colhead{NPS\_Field2} & \colhead{NPS\_Field3} & \colhead{NPS\_Field4} & \colhead{SPS\_Field1} }
\startdata 
{Standard Model}  & $694.99/341$ & $520.80/400$ & $412.17/370$ & $518.65/410$ & $520.39/505$ \\
{2T\_N\_Ne Model} & ${\bf 407.70/336}$  & ${\bf416.16/396}$  & ${\bf352.66/366}$  & ${\bf472.77/406}$  & ${\bf461.14/501}$  \\
{Model\_Willingale03} & $450.22/337$  & $460.92/397$  & $366.64/366$  & $485.21/406$  & $469.93/501$  \\
{Model\_Miller2008} & $446.71/336$  & $444.49/396$  & \nodata$^{a}$  & $478.77/405$  & $463.79/500$  \\
{1T\_ionized absorber} & $478.97/329$  & $456.01/396$  & $364.41/365$  & $471.11/405$  & $474.80/500$  \\
{Model\_NEI} & $562.02/340$  & $477.33/399$  & $435.35/369$  & $501.17/409$  & $528.53/504$  \\
\enddata
\tablecomments{\\
\footnotesize
Standard Model: One-temperature model with solar abundances.\\
2T\_N\_Ne Model: Neutral absorber two-temperature model with variable N and Ne.\\
Model\_Willingale03: Neutral absorber one-temperature model with variable O, Ne, Mg and Fe from Willingale et al. 2003.\\
Model\_Miller2008: Neutral absorber one-temperature model with variable N, O, Ne, Mg and Fe from Miller et al. 2008.\\
1T\_ionized absorber: Ionized absorber one-temperature model with variable O, Ne, Mg and Fe.\\
Model\_NEI: Neutral absorber non-equilibrium ionization model from Yamamoto et al. 2022.\\
$^{a}$ Model could not be constrained for this field.
}
\end{deluxetable*}

\clearpage

\bibliographystyle{aasjournal}

\end{document}